# On the Impact of Caching and a Model for Storage-Capacity Measurements for Energy Conservation in Asymmetrical Wireless Devices

Constandinos X. Mavromoustakis
Department of Computer Science,
University of Nicosia
46 Makedonitissas Avenue, P.O.Box 24005
1700 Nicosia, Cyprus
E-mail: mavromoustakis.c@unic.ac.cy

*Abstract:* **Traffic and channel-data rate combined with the stream oriented methodology can provide a scheme for offering optimized and guaranteed QoS. In this work a stream oriented modeled scheme is proposed based on each node's self-scheduling energy management. This scheme is taking into account the overall packet loss in order to form the optimal effective -for the end-to-end connection- throughput response. The scheme also -quantitatively- takes into account the asymmetrical nature of wireless links and the caching activity that is used for data revocation in the ad-hoc based connectivity scenario. Through the designed middleware and the architectural layering and through experimental simulation, the proposed energy-aware management scheme is thoroughly evaluated in order to meet the parameters' values where the optimal throughput response for each device/user is achieved.**

## I. INTRODUCTION

Today, QoS is undoubtedly one of the central pieces of the overall packet network technologies either being supported by a wireless interface or by a wired. Particularly when hosting delay sensitive applications in such networks, the offered QoS plays the major role for the end-to-end accurate communication (either form user's perspective or network's perspective).

The traffic that traverses the broadcast channels should interfere as a regulator for the self-tuning characteristics of each device in order to save energy and prolong the network lifetime. Network partitioning problems may occur if network interfaces are turned off for prolonged time, resulting in unacceptable QoS for end users. Hence, the energy conservation mechanism has to be closely collaborative with the incoming traffic characterization with the routing protocol behavior used by nodes, and with the time constrained behavior of the node being in the idle state. All the above should be considered in the forwarding decision performed by nodes while maintaining the packet forwarding mechanism. In this work this consideration has been done, in order to evaluate the impact of caching with the bounded model for storage-capacity measurements for energy conservation in such asymmetrical wireless links.

The organization of the paper is as follows: Section 2 discusses the related work that has been done on similar energy-aware conservation schemes and the conducted solutions by different schemes. Section 3 then introduces the proposed delay sensitive stream-oriented scheme and the impact of caching and storage-capacity measurements for energy aware QoS provision in asymmetrical wireless ad-hoc environments. Section 4 provides the evaluation of the proposed scheme and presents the simulation results focusing on the behavioral characteristics of the scheme and the newly introduced parameters. Finally section 5 concludes with a summary of our contribution and suggestions for further research.

## II. RELATED WORK

Today, the proposed schemes consider entirely symmetrical links with symmetric remaining capacity in the storage units. These schemes that are based only on protocols where different states are adopted randomly (switching node's states to "low energy"), cause severe degradation of network capacity [2]. There is already a significant amount of research work, which addresses the routing layer [7-10] for enabling energy conservation as well as the MAC and physical layers [11-15] by using different approximation techniques. This work neither involves any layered end-to-end mechanism nor enables any routing layer involvement. The proposed mechanism takes action on a MAC and physical layers and on application layer where the behavioral promiscuous caching is taking place, with respect to storage and capacity characteristics.

Many protocols have been designed and use different mechanisms to reduce energy consumption; they are classified into two categories: active and passive protocols. Active techniques conserve energy by performing energy aware conscious operations, such as scheduling the transmission slots using a directional antenna [16], and energy-aware routing [17]. The passive techniques conserve energy by scheduling network interface devices to the sleep mode when a node is not currently taking part in any communication activity (packet forwarding) or being the end recipient node.

Some of the recently proposed protocols deal with MAC layer [7] issues and network layer [17] issues and some are based on topological and geographical information-based techniques (GAF) [18]. Authors in [18] proposed a scheme based on the division of the entire network into small virtual pieces (grids). This area which is recognizable by geographical information, allows only one node to be active in the grid while the other nodes turn off their interfaces to conserve energy. In [19] the goal is to turn off nodes without significantly diminishing the capacity or connectivity of the network. Also a traffic-load history determination in association with battery lifetime has been examined in [4]. In [4] the research is focused on load history characterization for each node, targeting the energy conservation for delay and non-delay sensitive services. The self-similarity of packet traffic characterization studied in [4] allows nodes to change their state depending entirely on their traffic history.

As a part of the already existing work done in [1, 4] where a combination of history traffic scheme along with an association of the behavioral promiscuous caching characteristics and storage-capacity characteristics, this work extends the previous frameworks by measuring the validity and effectiveness of the proposed architectural middleware which controls the mobile interfaces and minimizes the energy consumed on any asymmetrically located wireless link. While some approaches based on overhearing or fading techniques do not actually address the association of *EC* (Energy Conservation) problem with any aspects of traffic, in this work this scenario is considered and examined. The channel's traffic is considered and it is combined with stream-traffic characteristics and the associated time that the caching mechanism wastes in order to temporarily save data packets destined for a node. The available capacity is then combined –based on the model in [1, 4]- with traffic and caching characteristics in order to enable minimization of the packet loss and end to end delays along with a way to conserve energy . An issue which is the research target is the association of the performance metrics with the asymmetrical characteristics of wireless devices like the capacity, the signal strength, and the channel load asymmetry (for loss sensitive traffic modeled and expressed in section III-(b)).

## III. PROMISCUOUS CACHING FOR EC BASED ON STREAM ORIENTED DELAY SENSITIVE APPROACH

### III-(a). A non-strict layering architecture

Nodes in wireless networks typically rely their survivability on their battery energy. Energy consumption should be minimized in order to save power for –initially- each wireless device independently or all of the WLANs' devices, in a decentralized form. The effects on reducing each node's power consumption has been studied thoroughly [4, 15-18]. The energy conservation mechanism has to be closely collaborative with the upper layer protocols used, to maintain the packet forwarding mechanism, in an error free mode. This work proposes a way to avoid any technical discrepancies that may exist between different devices, the energy consumption of each device or sudden energy deficiency of all the devices in a certain path.

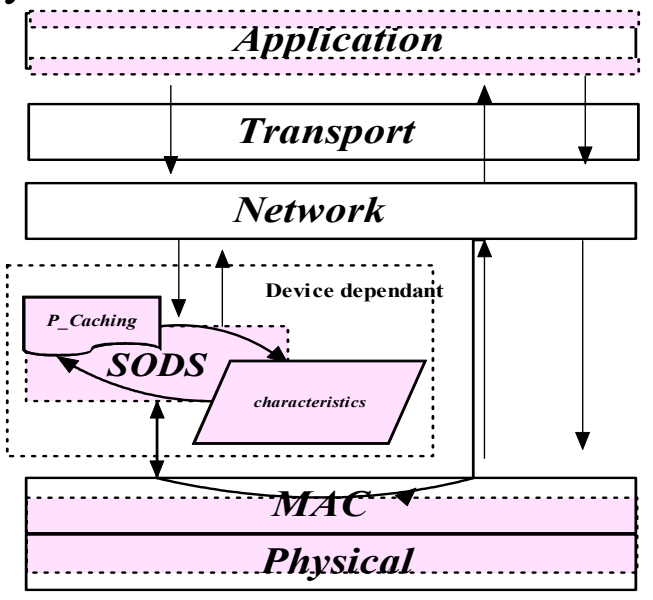

*Figure 1: Architectural communicational stack of Stream Oriented Delay Sensitive (SODS) mechanism which takes place in Physical, MAC layer and Application specific layers of the communication stack (non-strict layering structure).*

Figure 1 shows the main architectural communicational stack using the Stream Oriented Delay Sensitive (SODS) mechanism introduced in [1]. The SODS mechanism takes place in Physical, MAC and Application specific layers of the communication stack. Considering the variability of the distance covered by each device, the energy savings should be strictly device dependant. This distance dependant communication, can be associated with a particular caching-capacity metrics and can define a relation in power-capacity with respect to delay in power saving. Figure 1, addresses the combination of SODS mechanism with the effects of caching on each device independently. Each node validates the values (modeled and explained in the next section) and decides what is the next state of the node. Figure 1 shows in shaded form the SODS with caching along with the application layer collaboration. These collaborations among these layers are primarily responsible for all parameterized mechanisms that will be explained in next section.

### III-(b). Stream oriented approach for power consumption and the impact of capacity

According to traffic engineering, the energy that is consumed to transmit a data unit is directly proportional to the total energy consumption by a constant number. Thus if a sender wants to transmit a stream of data at rate R to a receiver, the corresponding transmission power P can be expressed as:

$$P = R \cdot d^r \text{, where } 2 < r < 4 \qquad (1)$$

In equation (1), $P$ is the consumed power, $r$ is the path loss exponent based on different channel models, $d$ is the distance between to adjacent wireless nodes, and R is data rate of the channel. Equation (1) assumes symmetrical channels in the communication between devices. In this case according to traffic engineering of symmetric links, power consumption over a long link is much higher than the total power consumption over several short links. Therefore:

$$(d_1 + d_2 + d_3 + ... + d_n)^r >> d_1^r + d_2^r + d_3^r + ... + d_n^r \quad (2)$$

Equation (2) is valid only if we consider symmetric links. Considering asymmetric links between nodes then the power consumed equals to:

$$P = \sum_{i=0}^{n} R_i \cdot d_i^r \quad (3)$$

where $R_i \cdot$ is the transmission rate of $i$ link, and $d_i^r$ is the distance of *i-node* to the next node-hop. It stands that:

$$d_1 \neq d_2 \neq d_3 \neq d_4 ... \neq d_n \quad (3.1)$$

In order to evaluate the lost packets we have first determined the ratio of transmitted and received blocks as follows:

$$RatioR = \frac{Recieved_Blocks}{Transmitted_Blocks} \quad (3.2)$$

Then the packet loss is measured as follows:

$$Packet_{loss} = 1 - (\frac{Recieved_Blocks}{Transmitted_Blocks}) \quad (3.3)$$

where the effective throughput $E_{ff}$ can be measured as follows:

$$EffectiveThroughput = E_{ff} = 1 - (Packet_{loss}) \cdot (\frac{PacketTrasferredSize}{PacketTrasferredTime}) \cdot (\frac{1}{Bandwidth}) \quad (3.4)$$

By using an exponential notation, we assume that power is reduced progressively [2, 4] with the remaining capacity on each node, which is evaluated by:

$$P_{i_c} = P_i \cdot e^{e^{C \cdot Eff_i}} \quad (4.1)$$

and

$$P_{i_C} = e^{e^{C_i \cdot Eff}} \sum_{i=0}^{n} R_i \cdot d_i \quad (4.2)$$

where C is the total data density. The associated caching eligibility of each node at any time during packet transfer is considered by using the following delay notation as indicated in [3]:

$$\overline{\delta}^{(m)} \approx \frac{\tau_0}{m} \log_2 i_P \quad (5.1)$$

where $m$ is the number of identical sized chunks that the file is divided, $i$ is the number of wireless peers in the path $P$ and $\tau_0$ is the amount of time taken to download the whole file, if downloaded from a single peer. Therefore a threshold for optimal caching (at any intermediate node) is chosen as:

$$\sigma_i = \frac{\sum_{i=0}^{n} R_i \cdot d_i}{\overline{\delta}^{(m)}}, \; \sigma_i \cdot \overline{\delta}^{(m)} = \sum_{i=0}^{n} R_i \cdot d_i \quad (5.2)$$

where $\sigma_i \cdot$ is the promiscuous caching threshold parameter or when minimized is considered as the optimal caching parameter for minimum energy consumption. The parameter $\overline{\delta}^{(m)}$ is the caching delay (duration in sec) as examined separately in [4]. It should also be noted that the shadowing and fading characteristics are considered as in [4] where there has been an association of the data delay during data revocation from a node, with the node fading characteristics $\Phi$ (as modeled in [4]). This estimation were taken into account with the combined ON-OFF periods (expressed in [1, 4, 20]) in order to efficiently reduce the active periods and thus to conserve energy.

Using the power estimation above along with the caching and capacity measurements and characteristics we considered an additional stream oriented approach as introduced in [1]. Packets are a part of streams which comprise a file –with file chunk correlations (like multimedia audio and video streams). All packets have a time $\tau$ for reaching destination in a free erroneous mode as expressed in [11] in order to arrive correctly at a specified destination. In our approach we also used the $S_n$ notation and streaming specification as used in [1]. $S_n$ is considered as the streaming packets parameter of a single application. These packets are marked as prioritized. The $S_{t-\tau,j}(S_1, S_2, S_3, ..., S_n)$ is called streaming delay bound, where j is the number of the possible intermediate nodes, that any of the stream packets $S_n$ might follow, and $S_{t-\tau,j}$ is the upper bound of the required time for correct reception of the stream, at the destination. The described scenario uses prioritized and non-prioritized packets (delay sensitive and "don't care" packets [5]). These packets have a bounded time delay $\tau$ to reach any specified destination. In our scheme "don't care" packets are further delayed onto intermediate nodes where prioritized packets are enforced to continue their "journey" to reach sooner their destination. Any packet delay is estimated by taking measures using the SODS mechanisms explored earlier, along with the bounded capacity and the caching threshold parameter [4] as a function of the effective throughput $E_{ff}$. Simulation results show that there is an optimization in the offered throughput along with the energy that every node consumes. This scheme saves energy while prolonging network's lifetime and offers an efficient scheme for optimizing the throughput of the system. Additionally since this scheme considers the probability of a missing block, it therefore takes into account the packet loss probability in the examination of all metrics mentioned above. It offers an in-time arrival of data packets as well as overall throughput optimization. The described methodology can significantly reduce the total consumed energy of all the terminals in the zone [16]. The proposed scheme provides upper protocol layer independency and enables the usage of the cross-layer feedback control mechanism through interfaces. This allows the lower level layers to adapt dynamically to changing network characteristics like capacity, energy and signal strength (distance).

## IV. PERFORMANCE ANALYSIS THROUGH SIMULATION AND DISCUSSION

### IV-(a). Specifications and routing protocol used

In the implementation of the proposed scenario the Zone Routing Protocol (ZRP) [16] is used. The specifications used for simulating our scheme are based theoretically on the WaveLAN PC/Card energy consumption characteristics found in the study by Feeney and Nilsson [22].

### IV-(b). Performance results of the proposed scenario

In our experiments we took into account the $\sigma_i$-the promiscuous caching threshold parameter (or bounded optimal caching parameter for minimum energy consumption) and values of $P_i$ in terms of $\Phi_i$ ( *Φ is the* fading channel characteristics parameter), for satisfying $Min(P_i,\Phi_i), Max(R_i), Min(\bar{\delta}^{(m)})$ mentioned in [4]. The promiscuous caching threshold parameter is estimated, taking into account many factors like end-to-end delay, hop-by-hop latency and power consumed over the delay and capacity measures. Some simulation experiments also were performed using different node capacities in order to evaluate the proposed scenario's response in contrast to node's required capacity for maintaining $Min(P_i,\Phi_i), Max(R_i), Min(\bar{\delta}^{(m)})$.

As previous researches cogently stated, the cached information destined for a proper node should be stored in a node with higher residual energy. As simulation process shows, if nodes with higher level of residual energy are chosen in the path, then the network partitioning probability [4, 5] is further reduced. Therefore the cached process which takes place is chosen on a recursive path basis [1] in order to face the discrepancies between packet delays and the storage capabilities of each node. The tree of Node Residual Energy (NRE) (expressed in [4, 15, 3]) is created in order to enable an on demand caching and to assign the certain packets to a certain node of a specified path.

For implementing the described scenario, we used the spine model of [4] (based on C/Objective C programming language). Topology of a 'grid' based network was modeled according to the grid approach described in [4]. In the simulation of the proposed scenario we used a two-dimensional network, consisting of 50 dense nodes. The topology changes dynamically as well as density and on a non-periodic basis (asynchronously as real time mobile users do). Each link (frequency channel) has max speed of 11Mb per sec (ideal speed), and the propagation path loss is the two-ray model without fading. The network traffic is modeled by generating constant bit rate (CBR) flows. Each source node transmits one 512-bytes (~4Kbits) packet. Packets are generated at every time step by following Pareto distribution as depicted in [4], and are destined for a random destination which is uniformly selected.

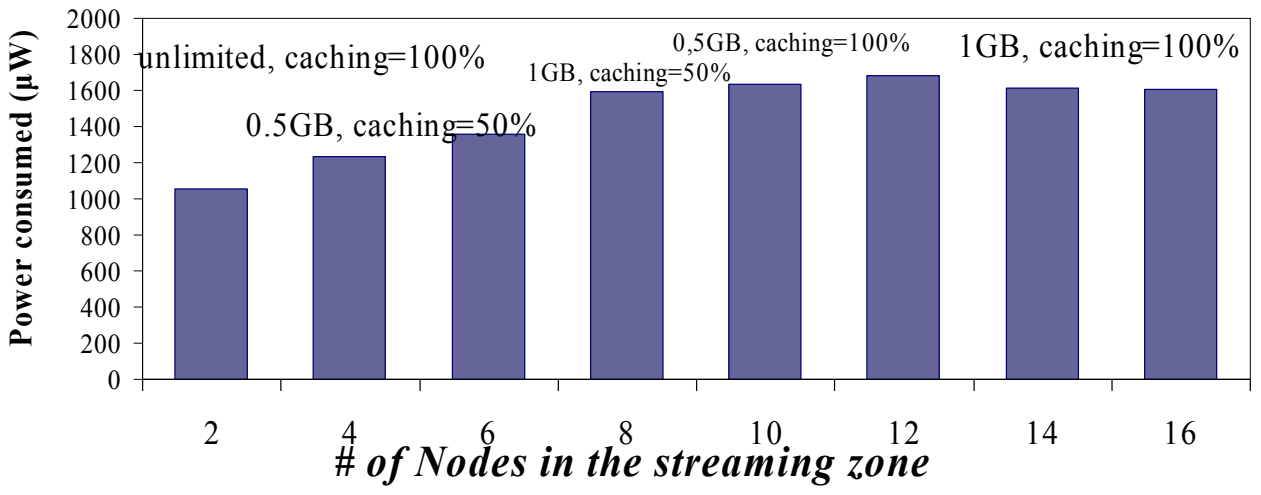

*Figure 5: The mean power consumed with the number of nodes being in the zone range.*

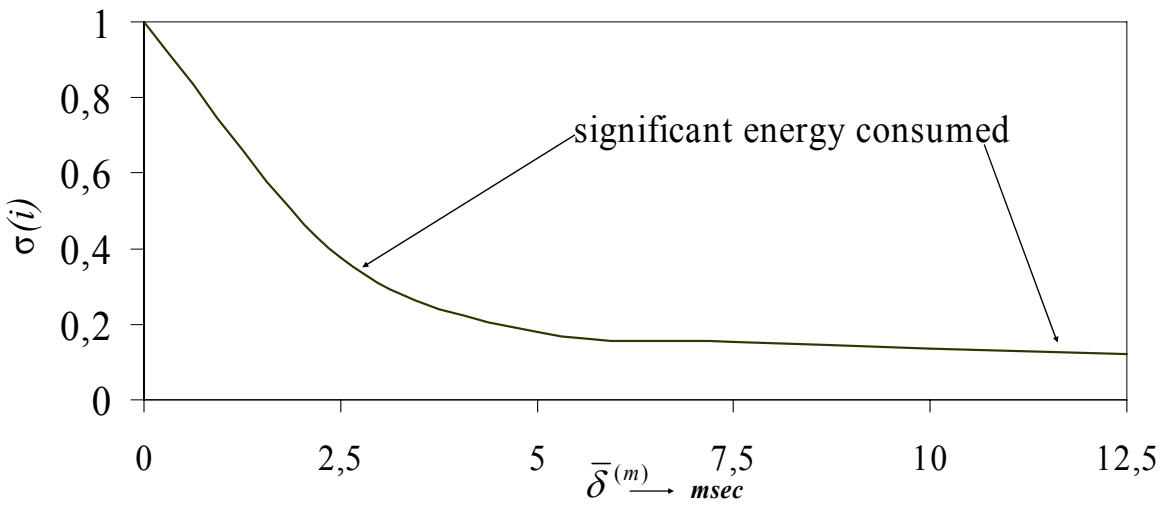

*Figure 6: The caching threshold parameter with the average caching delay experienced by all the involved mobile peers in the end-to-end forwarding activity.*

Figure 5 shows the mean power consumed with the number of nodes being in the zone range. It was found that for range distances less than 9m the power consumption of each device is kept at low level [4]. On the contrary by being higher than 9m then it increases dramatically. Figure 5 shows that by using the range coverage of the ZRP, the nodes that are set in the zone are considered to conserve higher energy. It is interesting to point out that even with 100% of caching activity and a 1GB total capacity, the energy consumed is kept in a desired upper bound/range (1602 μWatt). Figure 6 shows the promiscuous caching threshold parameter with respect to the average caching delay experienced by all the involved mobile peers during the end-to-end forwarding activity. This parameter is found to be ideal if is closer to the range of $0.2<\sigma<0.99$. If this threshold parameter is kept closer to 1 then the delay is strictly associated with the average caching delay experienced by mobile peers in the end-to-end forwarding activity as shown in Figure 6 taking into account the equation (4.2), (5.1) and (5.2). Additionally if fading channel intercalate with the above measures then the power of the radio waves decreases with the distance, and the power consumed is significantly higher.

Figure 7 shows the power consumed with respect to the average throughput. The proposed scheme can easily offer a throughput of 55% without consuming a significant amount of energy. It also of considerable interest that even a greater than >80% throughput is offered with less than 1100 μWatt being consumed. All measures were taken according to the scenario discussed earlier.

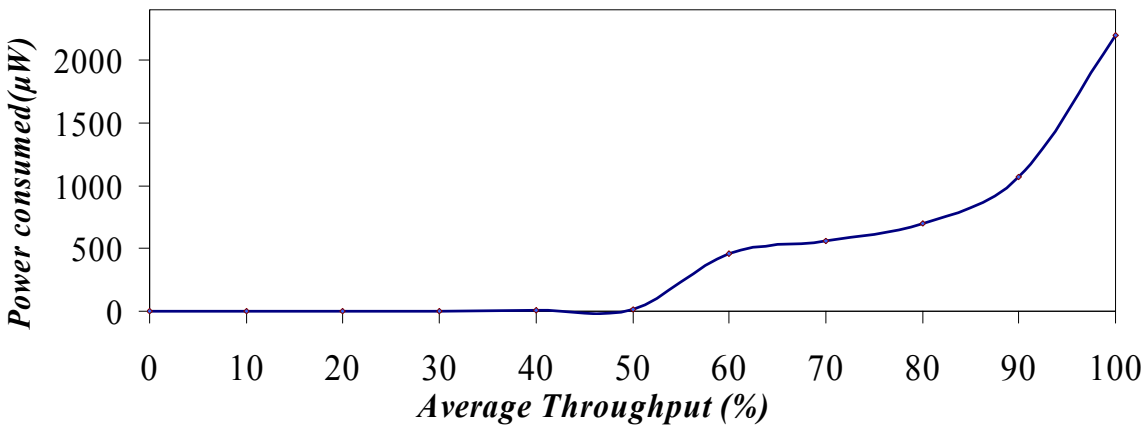

*Figure 7: The power consumed with respect to the average throughput.*

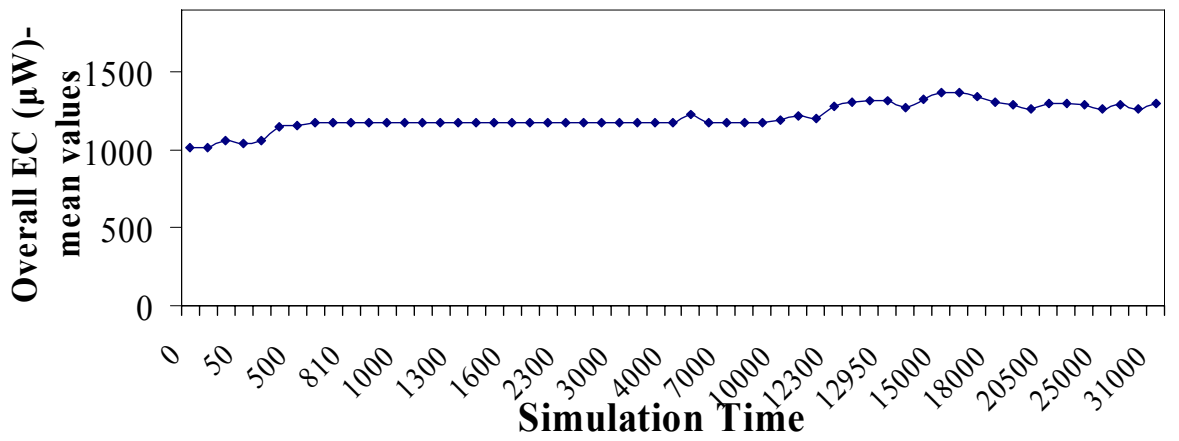

*Figure 8: Mean Energy consumption in each network's zone (ZRP-based) at any time during simulation.*

In figure 8 the mean Energy Consumption in each network's zone (using the zone of the ZRP) is illustrated, at any time during simulation. According to figure 8, it is shown that with proposed scheme the overall energy consumed is tuned. Overall, $P_i$ is significantly minimized satisfying the streaming delay bound and the parameters following the $Min(P_i, \Phi_i), Max(R_i), Min(\overline{\delta}^{(m)})$.

## V. CONCLUSIONS AND FURTHER RESEARCH

In this work, we have proposed a model for controlling the power consumption parameters and enable them into a bounded limit and a range of bounded values. The effective throughput is taken into account along with other metrics for the estimation of the energy consumed and the effective measure of the remaining capacity of the asymmetrically communicating links, connectivity, channel quality and fading characteristics, channels available bandwidth.

We are currently working towards an extension of this framework which takes into account other factors like network formation/partitioning problems as well as traffic engineering models in order to closely express such dynamically changing scenarios.